\documentclass[reprint,prb]{revtex4-1}

\usepackage{graphicx,epsf,epsfig}
\usepackage{subfigure}
\usepackage{dcolumn}
\usepackage{bm}
\usepackage{ulem}
\usepackage{hyperref}
\usepackage{lipsum}
\hypersetup{
    colorlinks,%
    citecolor=blue,%
    filecolor=blue,%
    linkcolor=blue,%
    urlcolor=blue
}
\begin{document}

\title{Quantum transport at the Dirac point: \\Mapping out the minimum conductivity from pristine to {disordered} graphene}
\author{Redwan N. Sajjad} 
\affiliation{Department of Electrical Engineering and Computer Science,
Massachusetts Institute of Technology, Cambridge, MA-02139, USA.}
\author{Frank Tseng} 
\affiliation{Naval Research Laboratory, Washington D.C. 20375, USA.}
\author{K. M. Masum Habib} 
\affiliation{Department of Electrical and Computer Engineering,
University of Virginia, VA 22904, USA.}
\author{Avik W. Ghosh}
\affiliation{Department of Electrical and Computer Engineering,
University of Virginia, VA 22904, USA.}
\date{\today}

\begin{abstract} 
The phase space for graphene's minimum conductivity $\sigma_\mathrm{min}$ is mapped out using Landauer theory modified for scattering using Fermi's Golden Rule, as well as the Non-Equilibrium
Green's Function (NEGF) simulation with a Monte Carlo sampling over impurity distributions. The resulting `fan diagram' spans the range from ballistic to diffusive over varying aspect ratios ($W/L$), and bears 
several surprises. {The device aspect ratio determines how much tunneling (between contacts) is allowed and becomes the dominant factor for the evolution of $\sigma_{min}$ from ballistic to diffusive regime. We find an increasing (for $W/L>1$) or decreasing ($W/L<1$) trend in $\sigma_{min}$ vs. impurity density, all converging around $128q^2/\pi^3h\sim 4q^2/h$ at the dirty limit}. In the diffusive limit, the {conductivity} quasi-saturates due to the precise cancellation between the increase in conducting modes from charge puddles vs the reduction in average transmission from scattering at the Dirac Point. In the clean ballistic limit, the calculated conductivity of the lowest mode shows a surprising absence of Fabry-P\'{e}rot oscillations, unlike other materials including bilayer graphene. We argue that the lack of oscillations even at low temperature is a signature of Klein tunneling.  
\end{abstract}

\maketitle


Since its discovery in the last decade, single layer graphene has catalyzed widespread research \cite{geim2007rise} stemming from its extraordinary material properties. Multiple
electronic, spintronic and opto-electronic applications are predicted to arise from the entire class of 2D materials emergent in graphene's footsteps \cite{fiori2014electronics}. Despite intense scrutiny, there exist many unresolved issues that continue to make the material fascinating. Among them is the physics of the minimum conductivity, $\sigma_\mathrm{min}$ around the Dirac point, where the density of states is expected to vanish. Instead of vanishing accordingly, $\sigma_\mathrm{min}$ for a  ballistic sheet with large width to length aspect ratio ($W/L \gg 1$) is shown to be a universal constant $\sigma_Q = {4q^2}/{\pi h}$ \cite{miao_07,twor_07}. 
\begin{figure}[ht!]
\includegraphics[width=3in]{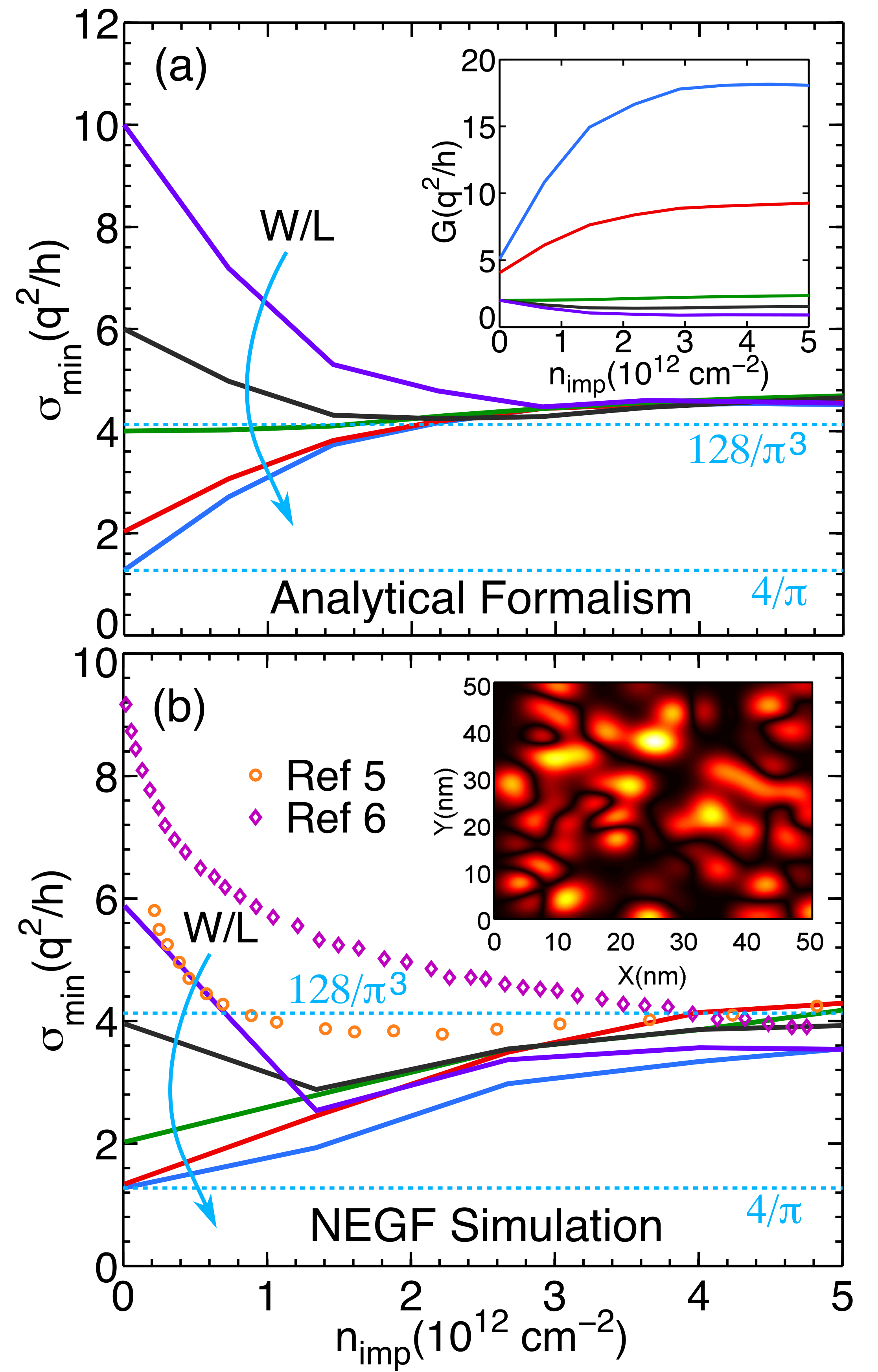}
\caption{ (Color online)(a) Fan diagram of quasi-analytical $\sigma_\mathrm{min}$ for $W=500$ nm with varying $W/L$ (inset shows conductance $G$). The ballistic $\sigma_\mathrm{min}$ is exactly at $\sigma_Q = {4q^2}/{\pi h}$. {{The two new features are (1) quasi-saturation at high impurity density to $\sim 128q^2/\pi^3h$ and (2) a flip in curvature between aspect ratios}}. (b) NEGF calculated $\sigma_\mathrm{min}$ averaged over puddle geometries (inset). The data saturate at $\sim 4q^2/h$ in dirty graphene. Dotted green curve is experimental data from Ref. \cite{chen_08} and open circles are theoretical predictions from Ref. \cite{adam_07} \label{phaseplot}}
\end{figure} 
This arises from the preponderance of tunneling through a continuum of subbands with near zero bandgaps. In these structures ($W>>L$ samples), a series of exponentially decaying tunnel transmissions adds up to an overall Ohmic term that factors out of the ballistic conductance $G = \sigma W/L$. Measured $\sigma_\mathrm{min}$s, however, are typically in the range $4-12q^2/h$ \cite{tan_07,chen_08,sui2011,amet2013}, except Ref. \cite{miao_07}, larger than $\sigma_Q$. This is surprising given that these experiments are mostly on dirty samples where we expect the conductivity to be not only non-universal, but certainly smaller than the ballistic limit. The increase in $\sigma_\mathrm{min}$ from $\sigma_Q$ arises from charged impurities on the substrates that create electron and hole puddles and contribute states to the charge neutrality point \cite{martin2007}. However an opposite, decreasing trend of $\sigma_\mathrm{min}$ vs. impurity concentration ($n_\mathrm{imp}$) was demonstrated theoretically by Adam $et. al.$ in Ref. \cite{adam_07} within Boltzmann transport theory, as well as experimentally in Ref. \cite{chen_08}. Clearly there are several disjointed pieces that have yet to
come together to provide a complete phase picture of the evolution of $\sigma_\mathrm{min}$ with sample quality.
\begin{figure}[htbp]
\begin{center}
\includegraphics[width=3.4in]{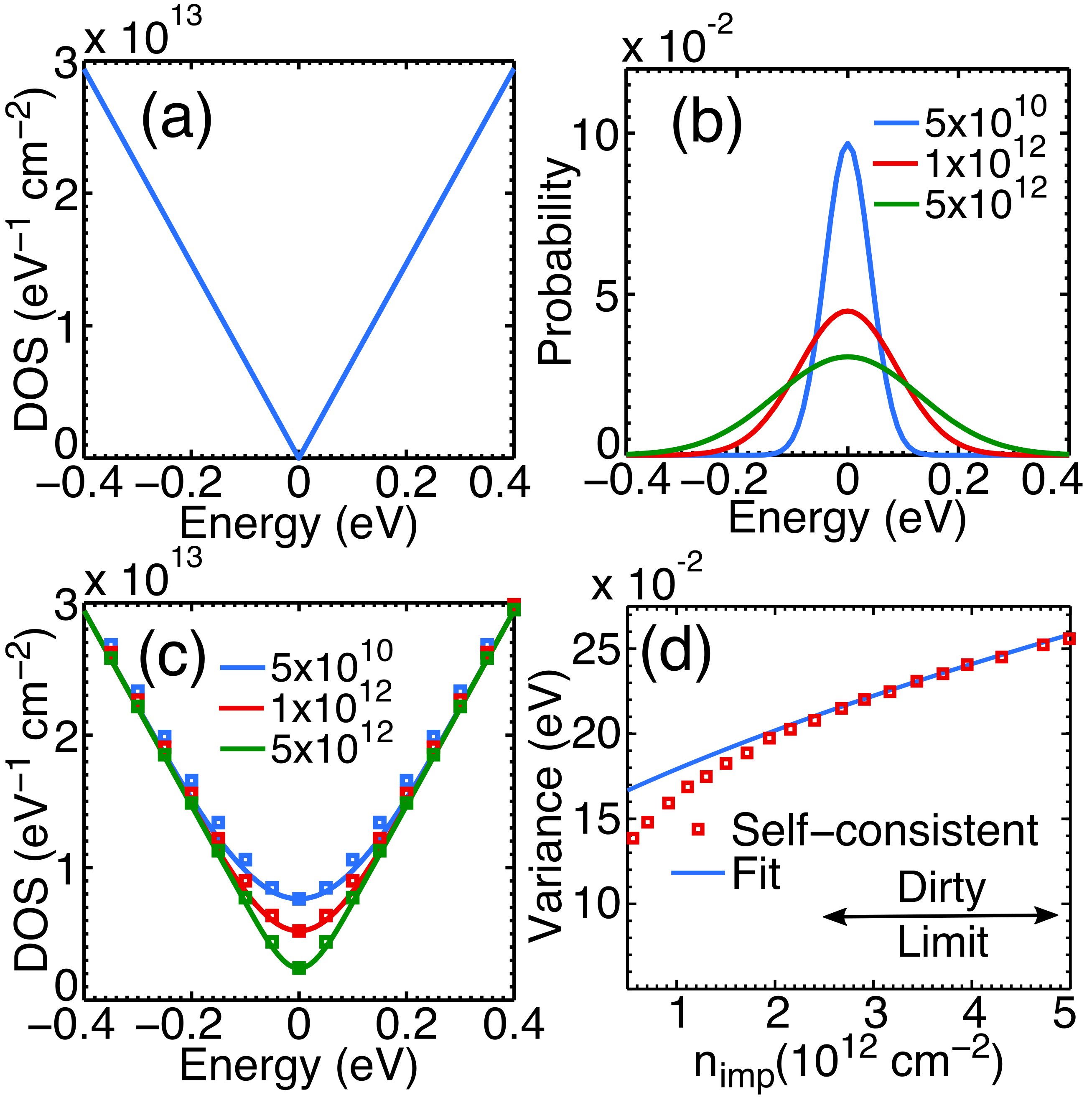}\caption{Averaging (a) the pristine graphene density of states with (b) a normal distribution of random potentials (c) erases the Dirac point. (d) The variance of the Gaussian is calculated self-consistently and refitted with a simplified expression, closely matching with the self-consistent calculation in the dirty limit.}
\label{AveDosPlot}
\end{center}
\end{figure}

In this paper, we use quasi-analytical Landauer equation as well as numerical NEGF (within the Fisher-Lee formulation) \cite{datta_05} to map out the entire
phase space of  $\sigma_\mathrm{min}$ for varying $n_\mathrm{imp}$ and $W/L$ (Fig.~\ref{phaseplot}). Our results clearly show that the missing link is the total tunneling current (a function of $W/L$), a piece of physics typically ignored in semi-classical models. The observed quasi-saturation arises due to a trade-off between the number of modes and the scattering time $\tau$ from charge puddles, as we move from the ballistic to diffusive regime. The total conductivity can be written as 
\begin{eqnarray}
\sigma = {G}_0\Big[M_pT_p+M_eT_e\Big] \times L/W
\label{sigtot}
\end{eqnarray} where ${G}_0 = {4}q^2/h$ is conductance quantum including spin and valley degeneracy, $M_p$ and $M_e$ are the number of propagating and evanescent modes and $T$ is the corresponding mode averaged transmission probability. While this equation defines an absolute lower bound on conductivity at $\sigma_Q = 4q^2/\pi h$ (dashed line in Fig. \ref{phaseplot} top), we will shortly show that for dirty samples with impurity density $\sim 3-5\times 10^{12}$/cm$^2$, it predicts a quasi-saturating $\sigma_\mathrm{min} \approx 4q^2/h$, consistent with experiments  (Fig. \ref{phaseplot}).
Part of the fan diagram for $W \ll L$, the decreasing trend in $\sigma_\mathrm{min}$ in Fig. \ref{phaseplot} obtained earlier using Boltzmann transport equation, arises naturally in our model from scattering of the propagating modes $\sigma \propto \mathcal{G}_0\Big[M_pT_p\Big]$, where $M_pT_p \propto \sqrt{n_0^2+n_\mathrm{imp}^2}/n_\mathrm{imp} = \sqrt{1+n_0^2/n_\mathrm{imp}^2}$ {($n_0$ is the background doping)}. For the opposite ballistic limit,  wide samples have a conductivity that dips down to the quantized value $\sigma_Q$ to generate the rest of the fan diagram. At the same time, narrow ballistic samples with limited tunneling show a {\it{conductance}} quantization ${\cal{G}}_0$ that bears a spectacular robustness with temperature and a remarkable absence of Fabry P\'{e}rot (FP) resonance even at low temperature. We interpret the absence of FP (Fig. 3) as a clear signature of Klein tunneling, where the linear relativistic electron transmits perfectly at normal incidence due to pseudospin conservation, contrary to the prediction of nonrelativistic Schr\"{o}dinger equation (which applies to Bilayer graphene as we show). Our results are supported by numerical NEGF sampled over a Monte Carlo distribution of charged impurities. 

\textit{Modeling charged impurities:} 
The lack of dangling bonds makes direct chemisorption of charged impurities difficult on graphene. However, dielectric substrates can have charged impurities that play a significant role on transport around graphene's Dirac point. The physisorption of charged impurities randomly dopes the graphene, creating a Gaussian distribution in energy of Dirac points around neutrality. The resulting erasure of the Dirac point is already seen in quantum capacitance measurements \cite{zhang2009origin}.
We can average the linear density of states of graphene (Fig.~\ref{AveDosPlot}a) over a Gaussian distribution of potentials (Fig.~\ref{AveDosPlot}b), with average potential zero, variance $\sigma_E$, and potential $U_i$ at the $i$th location.
The exact expression involving error functions was worked out by Li \textit{et. al.} \cite{li2011disorder}, but we can express it in a simpler form that interpolates between the low-energy parabolic and high energy linear behavior (Fig.~\ref{AveDosPlot}c). 
\begin{eqnarray}\label{doseq}
D_{\mathrm{puddle}}(E) &=& \frac{2\sqrt{\frac{2}{\pi}}\sigma_Ee^{-E^2/2\sigma_E^2} + 2|E|\mathrm{erf}(\frac{|E|}{\sigma_E\sqrt{2}})}{\pi\hbar^2v_F^2}\nonumber\\
D_\mathrm{puddle}&\approx& \frac{2\sqrt{E^{2}+2\sigma_{E}^{2}/\pi}}{\pi\hbar^{2}v_F^{2}}
\end{eqnarray} Eq.~\ref{doseq} shows that the variance $\sigma_E$ has a direct impact on the minimum density of states.  Fig.~\ref{AveDosPlot}d shows that $\sigma_E$ increases  with charged impurity concentration, so that {\it{the minimum number of modes for conduction is proportional to the statistical variance of charge impurities.}} This has also been worked out by solving Poisson's equation in cylindrical coordinates \cite{li2011disorder}
\begin{eqnarray}
	\sigma_{E}^{2}&=&2\pi n_\mathrm{imp}q^{2}\int[A_{k}]^{2}k\, dk  \label{eq:potVariance} \\
 	A_{k}&=&\frac{2e^{\displaystyle -\kappa z_{0}}Zq\, \mathrm{sinh}(k\, d)}{k\kappa_{ins}\mathrm{cosh}(k\, d)+(k\kappa_{v}+2\, q_{TF}\kappa)\mathrm{sinh}(k\, d)}  \label{eq:potA}   
 \end{eqnarray} $\kappa_v$ and $\kappa_{ins}$ are the respective vacuum and insulator dielectric constants, while $\kappa$ is their average. {And} $q_{TF}={2\pi q^{2}}/{\kappa}\,\,D_{\mathrm{puddle}}(E)$ is the Thomas-Fermi screening wave-vector which depends on the average density of states (Eq.~\ref{doseq}). $A_k$ is the potential solved from Poisson's equation which accounts for the distance of the impurities ($z_o$) inside the oxide, thickness of the oxide ($d$) and the screening length ($1/q_{TF}$). Solved self-consistently (between $\sigma_E$ and $D_{\mathrm{puddle}}$) we determine the variance of the normal distribution of potentials (Fig. \ref{AveDosPlot}d). Over the dirty {range}, we can simplify it with a fitted equation 
\begin{equation}
\sigma_E^2 \approx 2\hbar^2v_F^2n_\mathrm{imp}+C
\label{sigsimp}
\end{equation}where $C=0.027$eV$^2$. This equation closely approximates the self-consistent calculation at the dirty limit. The variation of $\sigma_\mathrm{min}$ in presence of charged impurities allows us to quantify the competition between increasing modes and increased scattering. 

\textit{Analytical Model for $\sigma_\mathrm{min}$.} The Landauer conductivity intuitively  frames conduction as proportional to the transmission probability of electrons, $T_n$, summed over all propagating and evanescent modes, where $n$ is the mode index.  $\sigma_\mathrm{min} = G\,{L}/{W}={4}q^{2}/{h}\sum\limits_{n=0}^{\infty}{T_n}{L}/{W}$. The general form for $T_n$, derivable by matching the pseudospinor wavefunctions across an $n$-$p$-$n$ or $p$-$n$-$p$ junction with barrier height $U_0$ gives \cite{twor_06} $T_{n}=\left|\frac{k_n}{k_n\cos{k_nL}+i(U_o/\hbar v_F)\sin{k_nL}}\right|^{2}$ where 
$k_n=\sqrt{(U_{o}/\hbar v_F)^{2}-q_{n}^{2}}$ {and $q_n = n\pi/W$
is the transverse wave-vector} in the channel that we sum over to get
the total transmission. When $k_n$ is real then the transverse modes
are propagating, while when $k_n$ is imaginary they become
evanescent. Imaginary $k_n$ changes all the trigonometric functions
to hyperbolic functions giving us an evanescent transmission {$T_e = 1/\cosh^2{q_nL}$} when $U_{o}$ is zero. An integral over a continuum of such cosh contributions
gives an overall factor of {$W/\pi L$} which leads to the ballistic conductivity quantization ($\sigma_Q$). 
For propagating modes, the transmission probability $T_p$ picks up an additional {scattering coefficient} term from a series sum over the multiple scattering history, $\lambda/(\lambda +L)$, where $\lambda$ is the electron mean free path in the presence of embedded impurities. The mean free path is $v_F\tau_{sc}$ where the momentum scattering time $\tau_{sc}$ is determined from Fermi's Golden Rule below.
Combining all the elements in Eq. \ref{sigtot}, we
arrive at the fan diagram in Fig. \ref{phaseplot}.

Impurity scattering occurs through a 2D screened Coulomb energy, given at long wavelength by the Thomas Fermi equation, $V_C({\bf{r}}) = {q^2}/({4\pi\epsilon_0r})e^{-\kappa r}$.
Using the pseudospin eigenstates, $\Psi_{i,f}({\bf{r}}) = 1/\sqrt{2S}\left(1~~~e^{i \theta_{i,f}}\right)^{T}
    e^{i\bf{k_i}.{r}}$ normalized over area $S$, we calculate the scattering matrix element $V_{if} = \int {d^2}{\bf{r}}\displaystyle\Psi^*_{f}({\bf{r}})V_C({\bf{r}})\Psi_{i}({\bf{r}})$. In terms of scattering wavevector and angle  $\text{$\Delta\bf{k} = \bf{k}_f-\bf{k}_i$,~~$\Delta\theta = \theta_f-\theta_i$}$, 
\begin{eqnarray}
V_{if}&=& \displaystyle \frac{q^2}{4\epsilon_0S\sqrt{\Delta k^2+\kappa^2}}\left[1+e^{\displaystyle i\Delta\theta}\right]
\end{eqnarray}
We can change to energy variables for elastic scattering using $|{\bf{k}_f}| = |{\bf{k}_i}| = E/(\hbar v_F),
(\Delta k)^2 = |{\bf{k}_f}-{\bf{k}_i}|^2 = {k_f^2 + k_i^2 -2k_fk_i\cos{\Delta\theta}}= 2E^2(1-\cos{\Delta\theta})/(\hbar^2 v_F^2)$.
For an impurity density $n_\mathrm{imp}$ and cross sectional area $S$ (i.e., number of impurities $n_\mathrm{imp}S$), Fermi's Golden rule now gives us, $\hbar/\tau_{sc} = \sum_{{\bf{f}}}|V_{if}|^2\delta(E-E_k)(1-\cos\theta_k){n_\mathrm{imp}}S$. Converting sum into integral using the density of states (Eq.~\ref{doseq}), and using the calculated expression for $|V_{if}|^2$ simplified for low energies, we get
\begin{eqnarray}
\frac{\hbar}{\tau_{sc}} =  \frac{q^4\hbar^2v_F^2n_\mathrm{imp}}{16\epsilon_0^2\pi}
\int D_\mathrm{puddle}(E_k)dE_k\delta(E-E_k)\nonumber\\\times\int d\Delta\theta\frac{1-\cos^2\Delta\theta}{2E_k^2(1-\cos\Delta\theta) + \hbar^2v_F^2\kappa^2}
\end{eqnarray}
The cosine integral followed by the delta function energy integral gives us 
\begin{eqnarray}
\frac{\hbar}{\tau_{sc}} =  \frac{q^4\hbar^2v_F^2n_\mathrm{imp}}{16\epsilon_0^2\pi}
D_\mathrm{puddle}(E)
\frac{\pi}{2E^4}\nonumber\\\times\left[2E^2+\hbar^2v_F^2\kappa^2 - \hbar v_F\kappa
\sqrt{4E^2+\hbar^2v_F^2\kappa^2}\right]
\end{eqnarray}
with $D_\mathrm{puddle}$ defined in Eq. \ref{doseq}. For $E \ll \hbar v_F\kappa$, the term in square brackets
expands to $ 2E^4/\hbar^2v_F^2\kappa^2+ O(E^6/\hbar^4v_F^4\kappa^4)$. We then get
\begin{equation}
\displaystyle\frac{\hbar}{\tau_{sc}} \approx  \frac{q^4n_\mathrm{imp}D_\mathrm{puddle}}{16\epsilon_0^2\kappa^2}
\end{equation}
with $\kappa = q^2D_\mathrm{puddle}/\epsilon_0$, giving us ${\hbar}/{\tau_{sc}} =  {(n_\mathrm{imp}}/{16)D_\mathrm{puddle}}$.
Using the Einstein relation (diffusion coefficient ${\cal{D}} = {v_{F}^2\tau_{sc}}/{2}$), we get
\begin{equation}
\sigma_\mathrm{min} = q^2D_\mathrm{puddle}{\displaystyle\cal{D}} = \displaystyle\frac{8 q^2v_F^2\hbar}{n_\mathrm{imp}}D_\mathrm{puddle}^2
\end{equation}
At high impurity density, $D_\mathrm{puddle}^2 \approx 8\sigma_E^2/\pi^3\hbar^4v_F^4$ (Eq. \ref{doseq}). Using the approximate relation from Eq.~\ref{sigsimp} matching the self-consistent calculation fairly well in the dirty limit (Fig. \ref{AveDosPlot}), we get 
\begin{equation}
\lim_{\displaystyle n_\mathrm{imp} \rightarrow \infty}\sigma_\mathrm{min} \approx \displaystyle\frac{128 q^2}{\pi^3 h} = 4.12\frac{q^2}{h} 
\end{equation} 

\textit{Numerical model for $\sigma_\mathrm{min}$}: 
{We now show NEGF based numerical simulation results to calculate $\sigma_{\mathrm{min}}$ in presence of charged impurities. We implement a discretized $\textbf{k.p}$ Hamiltonian ($H$) to expedite computation.} We use a sequence of Gaussian potential profiles for the impurity scattering centers, 
\begin{eqnarray}\label{gauss}
U(r) = \sum_{n=1}^{n_\mathrm{imp}} U_n\exp{(-{|r-r_n|^2}/{2\zeta^2})}
\end{eqnarray} specifying the strength of the impurity potential at atomic site $r$, with ${r_n}$ being the positions of the impurity atoms and $\zeta$ the screening length ($\sim$3 nm). The amplitudes $U_n$ are random numbers following a Gaussian distribution with a standard deviation of 100meV \cite{sui2011}. {This standard deviation is to be differentiated from the standard deviation in the density of states description (Eq. \ref{doseq})}, which is a lumped description for the entire sheet instead of individual impurities. {The Gaussian profile (Eq. \ref{gauss}) is used to prevent the potential from going to infinity at the scattering centers (Thomas-Fermi) and such approach is widely employed in the literature \cite{klos_10, lewenkopf_08,adam_09,sui2011,rycerz2007}}.
With $U$ added to $H$, we calculate $\sigma_{min}$ as a function $n_\mathrm{imp}$ (Fig. \ref{phaseplot}b) by calculating average conductance over $\sim$800 random impurity configurations. In the ballistic limit, $\sigma_\mathrm{min}$ varies linearly with L/W, but as the sample gets dirtier, the $\sigma_\mathrm{min}$ becomes less dependent on L/W. At high impurity limit, $\sigma_\mathrm{min}$ becomes weekly dependent on $n_\mathrm{imp}$ and saturates around 4$q^2/h$. In most experiments, the device length $L$ is larger than width $W$ and therfore see a decreasing trend for $\sigma_\mathrm{min}$ vs. $n_\mathrm{imp}$ such as Ref.~\cite{chen_08}. The evolution of $\sigma_\mathrm{min}$ from ${4q^2}/{\pi h}$ to $\sim 4q^2/h$ and therefore the missing $\pi$ can only be seen for devices with $W>>L$. The differences between the numerical and the analytical approaches most likely originate from the lack of adequate samples.  

\begin{figure}
\centering
\includegraphics[width=3.4in]{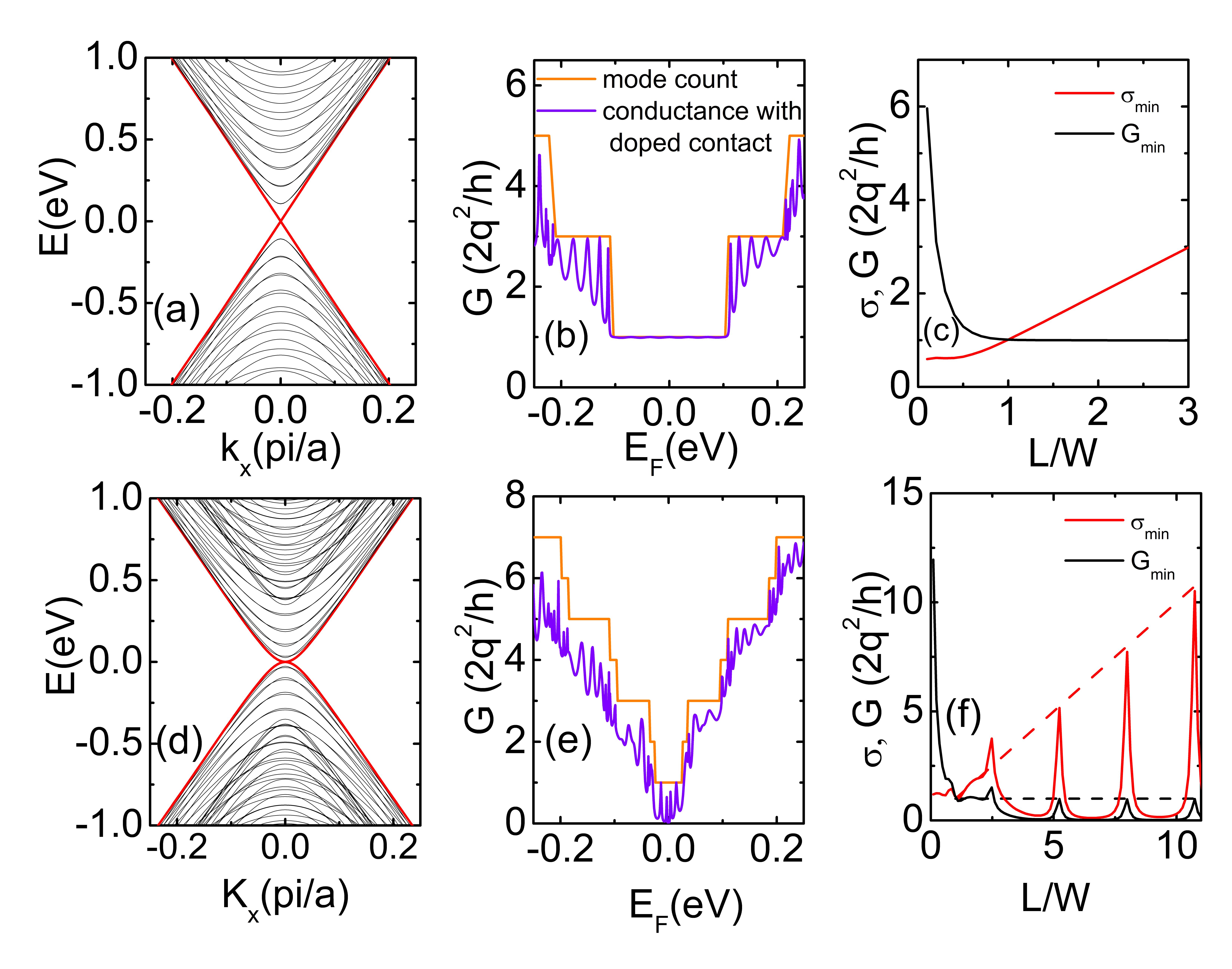}\quad
\caption{(Color online) NEGF calculation of total conductance $G$ of single layer graphene and bilayer graphene reveals {the nature of Fabry-P\'{e}rot oscillation for the lowest mode. a,d shows linear and parabolic $E-K$ in single layer and bilayer graphene. The lowest mode in single layer does not show any oscillation (b) but the bilayer does (e)}. The variation of minimum conductance and conductivity for single layer shows saturating $G_\mathrm{min}$ at $2q^2/h$ (c), while for bilayer graphene the minimum conductance never saturates and produces oscillation in both $G_\mathrm{min}$ and $\sigma_\mathrm{min}$ (f).}
\label{fig3}
\end{figure}
%
\textit{Absence of Fabry-P\'{e}rot as a signature of Klein tunneling:}~
Due to non-uniform doping along the metal-graphene-metal captured in our model by the differential dopings, a Fabry-P\'{e}rot cavity is formed. Such a cavity leads to quantum interference oscillations and conductance asymmetry ($n$-$n$-$n$ vs. $n$-$p$-$n$ doping), seen in Fig.~\ref{fig3} in the ballistic limit. Such
oscillations have been seen experimentally at low temperature in 2DEGs \cite{beenakker1991quantum}, but are conspicuously missing for the
lowest mode in single layer graphene (SLG), as seen in Fig.~\ref{fig3} left column. In contrast, the higher modes in the same column show oscillations, as do all the modes for bilayer graphene (BLG) seen in the right column. The lowest mode in single layer graphene has forward and reverse propagating $E-k$ bands with opposite pseudospin indices (bonding vs antibonding combinations of dimer $p_z$ orbitals) that disallow any reflection at heterojunctions. The resulting Klein tunneling \cite{katsnelson_06} makes the heterojunctions completely transparent to the lowest propagating modes and eliminates any Fabry-P\'{e}rot oscillations. The parabolic lowest bands of BLG have twice the winding number around the Fermi circle (angle $2\theta_{i,f}$ in the pseudospin eigenstate $\Psi_{i,f}$) and thus a common pseudospin index, leading to finite reflection and Fabry-P\'{e}rot oscillations. 

We thus expect distinct behaviors of $\sigma_\mathrm{min}$ vs. $L/W$ in single layer and bilayer graphene. For large $L/W$, $G_\mathrm{min}$ for SLG approaches $2q^2/h$ eliminating all tunneling modes from source to drain and $\sigma_\mathrm{min} = GL/W$ increases linearly (Fig. \ref{fig3}(a-c), already demonstrated in experiment \cite{miao_07}. 
For BLG (Fig. \ref{fig3}d-f), the conductance oscillation for the lowest mode is manifested in the length dependence as well, leading to an oscillation in both $G_\mathrm{min}$ and $\sigma_\mathrm{min}$. For small $L/W$, the $\sigma_\mathrm{min}$ saturates to $4q^2/(\pi h)$ and $2q^2/h$ for SLG and BLG respectively \cite{twor_06,katsnelson_ejp}. Such nontrivial transport behavior near the Dirac point is a measurable signature of Klein tunnel and reflection. 

\textit{Conclusion:}
The composite phase plot of graphene's minimum conductivity is presented within {a unified Landauer-Fermi's golden rule and NEGF} transport model. We show a general convergence of $\sigma_\mathrm{min}$ vs. impurity concentration along with a quasi-saturation at high impurity concentration to $\sim 4q^2/h$ irrespective of device dimensions. For high aspect ratios the increase in density of states due to charged impurities results in a logarithmically increasing $\sigma_{min}$ from the ballistic limit. On the other hand for low aspect ratios, the scattering due to charged impurities dominates and results in a power law decrease in the $\sigma_{min}$. For clean samples with conductance quantization, gating the sample into its lowest mode reveals a striking absence of low-temperature Fabry-P\'{e}rot oscillations at low temperatures for SLG but not BLG, providing a signature of Klein tunneling.

\textit{Acknowledgement:} This work was financially supported by the NRI-INDEX center. The authors thank Eugene Kolomeisky (UVa) for useful discussions.
\bibliographystyle{apsrev4-1}
\bibliography{sajjad_jab,sajjad_library2}

\end{document}